\documentclass[amsmath,amssymb,aps,prd,nofootinbib,showkeys,reprint, floatfix,groupedaddress,superscriptaddress,amsmath,amssymb]{revtex4-1}

\usepackage[breaklinks = true,colorlinks = true,linkcolor = blue,citecolor = blue]{hyperref}
\usepackage{graphicx}

\newcommand{\Nc}{N_{\text{c}}}
\newcommand{\fm}{\,\text{fm}}
\newcommand{\MeV}{\,\text{MeV}}

\newcommand{\br}{\boldsymbol{r}}
\newcommand{\btau}{\boldsymbol{\tau}}
\newcommand{\calO}{\mathcal{O}}
\newcommand{\rhard}{r_{\text{hard}}}
\newcommand{\rsoft}{r_{\text{soft}}}
\newcommand{\rhoB}{\rho_{\text{B}}}
\newcommand{\rhoS}{\rho_{\text{S}}}
\newcommand{\pc}{p_{\text{c}}}
\newcommand{\pq}{p_{\text{q}}}

\begin{document}

\title{Hard-core deconfinement and soft-surface delocalization from nuclear to quark matter}

\author{Kenji Fukushima}
\email{fuku@nt.phys.s.u-tokyo.ac.jp}
\affiliation{Department of Physics, The University of Tokyo, 
  7-3-1 Hongo, Bunkyo-ku, Tokyo 113-0033, Japan}

\author{Toru Kojo}
\email{kojo.toru@gmail.com}
\affiliation{Key Laboratory of Quark and Lepton Physics (MOE) and
  Institute of Particle Physics, Central China Normal University,
  Wuhan 430079, China}

\author{Wolfram Weise}
\email{weise@tum.de}
\affiliation{Physics Department, Technical University of Munich, 85748
Garching, Germany}

\begin{abstract}
  We propose a conceptual distinction between hard and soft
  realizations of deconfinement from nuclear to quark matter.  In the
  high density region of Hard Deconfinement the repulsive hard cores
  of baryons overlap each other and bulk thermodynamics is dominated
  by the core properties that can be experimentally accessed in
  high-energy scattering experiments.  We find that the equation of
  state estimated from a single baryon core is fairly consistent with
  those empirically known from neutron star phenomenology.  We next
  discuss a novel concept of Soft Deconfinement, characterized by
  quantum percolation of quark wave-functions, at densities lower than
  the threshold for Hard Deconfinement.  We make a brief review of
  quantum percolation in the context of nuclear and quark matter and
  illustrate a possible scenario of quark deconfinement at high baryon
  densities.
\end{abstract}
\maketitle

\section{Introduction}

Microscopic mechanisms of the deconfinement phenomenon from nuclear to
quark matter are still veiled in mystery.  There are countless numbers
of theoretical attempts since the first speculation on quark matter in
Refs.~\cite{Itoh:1970uw,Collins:1974ky}, but we have not established
any convincing picture to understand deconfinement along the baryon
density axis (see
Refs.~\cite{Fukushima:2010bq,Fukushima:2013rx,Buballa:2014tba,Fischer:2018sdj}
for reviews on the phase diagram and various conceivable scenarios at
high density, see also Ref.~\cite{Luo:2017faz} for a review on
experimental efforts to reveal the phase diagram).  To gain a
theoretical understanding, solving the first-principles theory, i.e.,
Quantum Chromodynamics (QCD) is not necessarily a unique route as it
is severely hindered by the notorious sign
problem~\cite{Muroya:2003qs,Aarts:2015tyj}.  In this work we are
proposing a novel viewpoint to make a conceptual differentiation
between hard and soft realizations of deconfinement and interpret
quark matter based on a condensed-matter physics analogue.

To make our proposal clearer, it is instructive to start with our
understanding of deconfinement in a different environment,
\textit{at high temperature $T$ and low baryon density $\rho$}, rather
than high density and low temperature of our current interest.  The
great advantage in such an environment at high $T$ and low $\rho$ is
that the lattice-QCD simulations provide detailed quantitative
information on deconfinement along the $T$-axis as long as $\rho$ is
small.  It is known that deconfinement from hadronic matter to a
quark-gluon plasma (QGP) occurs continuously at physical quark
masses.  In other words, there is no genuine phase transition in a
strict sense: physical degrees of freedom should be smoothly connected
from hadronic to QGP matter.  We note that there are several
approximate and phenomenological measures of quark deconfinement by
means of fluctuations~\cite{Asakawa:2000wh,Jeon:2000wg}.  Especially
the quartic to quadratic ratio of the baryon number fluctuation was
first considered in this way~\cite{Ejiri:2005wq}.  The lattice-QCD
results including these fluctuations gave us a guiding principle, but
it is still nontrivial whether gauge-invariant thermodynamic
quantities can diagnose contents of physical degrees of freedom.
They should be hadrons at low $T$ and change to quarks and gluons at
high enough $T$, but these degrees of freedom should be converted in
an intermediate region without a phase transition.

The success of the hadron resonance gas (HRG) model has been a
milestone for studies on
deconfinement~\cite{BraunMunzinger:1994xr,Cleymans:1996cd,Cleymans:1999st,BraunMunzinger:2001ip,Becattini:2005xt,Andronic:2005yp}.
Actually, in the large-$\Nc$ limit (where $\Nc$ denotes the number of
colors), mesons are noninteracting particles and the deconfinement
phenomenon of meson-dominated matter is essentially a Hagedorn
transition of
mesons~\cite{Cabibbo:1975ig,Andronic:2009gj,Cohen:2011yx}, leading to
blow-up behavior of thermodynamic quantities.  Later on, it has been
demonstrated that the excluded volume effect would tame singular
behavior of thermodynamic quantities~\cite{Andronic:2012ut}, which
indicates that such a transient state from hadronic matter to the QGP
could be approximated well by interacting mesons.  This suggests that
inter-hadronic interactions are gradually taken over by partonic
degrees of freedom, i.e., quarks and gluons.  Here, let us emphasize
two properties of high-$T$ deconfinement:
\begin{enumerate}
\item
  The fact that there is no sharp phase transition allows for
  significant overlaps of the hadronic and the partonic regimes.  The
  duality in thermodynamics between interacting hadrons and partons
  holds at overlapping densities.  The Polyakov loop is only an
  approximate order parameter and there is no strict order parameter
  for deconfinement in the presence of dynamical quarks.  One may
  think that the color conductivity (which has been computed in weakly
  coupled theories~\cite{Heiselberg:1994px,Arnold:1998cy}) might play
  the role of an order parameter, but its behavior is expected to be
  smooth similarly to the Polyakov loop.
     
\item
  Although there is no clear-cut separation, if thermodynamic
  quantities are dominated by hadrons (or partons), it would be
  reasonable to call such a state the hadronic phase (or the QGP,
  respectively).  In the hadronic phase the color-singletness should
  be imposed locally, but at high enough $T$ more and more multiple
  interactions would be involved and gradual deconfinement can be
  possible effectively even under the condition of color neutrality.
\end{enumerate}

One might think that the intuitive understanding based on the HRG
model could apply to the high density matter as well, but this
expectation would not work straightforwardly.  The crucial difference
is manifest especially in the large-$\Nc$ limit:  nucleons are heavy
and their kinetic energy is suppressed by $\Nc$, while their
interaction energy is enhanced as $\calO(\Nc)$~\cite{Witten:1979kh}.
Given such strong interactions, it is far from trivial whether
nucleons are really the relevant degrees of freedom or not.  In other
words, we must revise our notion of deconfinement due to interaction
effects when dense baryonic matter is concerned.

The invention of Quarkyonic Matter~\cite{McLerran:2007qj} has invoked
a lot of theoretical arguments along these lines.  Quarkyonic Matter
refers to baryonic matter whose pressure is $\calO(\Nc)$ from the
enhanced interaction energy.  At a microscopic level baryonic
interactions originate from $\Nc$ permutations of color-singlet quark
exchanges, and so one may well consider that this $\calO(\Nc)$
pressure of baryonic matter is dominated by quarks even in the
confined phase in which excitations on top of the Fermi surface should
still be baryons.  In this theoretically idealized world with
$\Nc\to\infty$ the picture of Quarkyonic Matter is well-defined.  It
is important to note that not only two-body but all many-body
color-singlet interactions in Quarkyonic Matter are of the same order
$\sim \calO(\Nc)$ according to the large-$\Nc$
counting~\cite{Witten:1979kh}.

Now, we must emphasize the importance of departing from the idealized
setup and refine the deconfinement picture for $\Nc < \infty$.  The
idea of Quarkyonic Matter has posed an interesting question that was
considered long ago in nuclear physics:  the short-range repulsive
hard-core interaction between nucleons can be viewed as arising from
quark exchanges exactly in the same way as in the Quarkyonic Matter
argument, while the long-range interaction via meson exchanges also
involves quark-antiquark exchange mechanisms at a microscopic level
\footnote{In this context pions are special because of their role as chiral Nambu-Goldstone bosons.}.
So, on the one hand, there appears to be no principal difference
between these two types of interactions.  On the other hand, as we
shall point out, the separation between short- and long-distance
nuclear interactions has its correspondence in a delineation of hard
and soft scales in the structure of the nucleon itself, and this is
the baseline for our subsequent discussion of ``hard'' and ``soft''
deconfinement.

\begin{figure}
  \includegraphics[width=0.4\columnwidth]{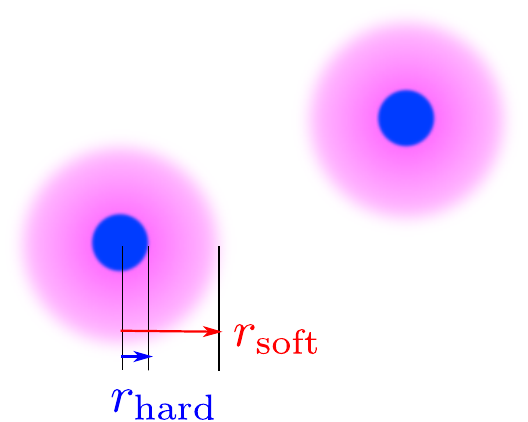}
  \caption{Nucleons are characterized by two scales, $\rhard$
    representing the hard core radius, and
    $\rsoft$ representing the size of the surrounding meson clouds.}
  \label{fig:dilute}
\end{figure}

Let us visualize our Gedankenexperiment of deconfinement.  Nucleons at
low energy can be viewed approximately as composed of a hard core
surrounded by meson (or quark-antiquark pair) clouds as picturized in
Fig.~\ref{fig:dilute}.  Typically, the core radius is around
$\sim 0.5\fm$, while the radius of meson clouds is around $\sim 1\fm$
as will be substantiated further in the next section.  If the baryon
density is so high that the hard cores begin to overlap, quark matter
is unambiguously realized.  We shall call this
\textit{Hard Deconfinement} of quark matter.  Once Hard Deconfinement
occurs, we can infer quantitative properties of such dense quark
matter from the internal nucleon structures, as we will demonstrate
for the construction of the equation of state (EoS) in the present
work.

One may also identify Hard Deconfinement with ``valence'' quark
deconfinement.  The baryon number distribution in a single nucleon is
localized in the core region: the $\Nc$ valence quarks in the core add
up to the net baryon number.  In this way, especially in the
large-$\Nc$ limit, Hard Deconfinement can be clearly defined in terms
of baryon transport:  in the confined phase the baryon number
transport is suppressed by heavy baryon masses, while quarks can
transport the baryon number once Hard Deconfinement occurs.  Meson
clouds, on the other hand, carry no net baryon number.  The baryon
number conductivity can be interpreted as the heat conductivity which
can therefore act as an approximate order parameter for Hard
Deconfinement.

\begin{figure}
  \includegraphics[width=0.4\columnwidth]{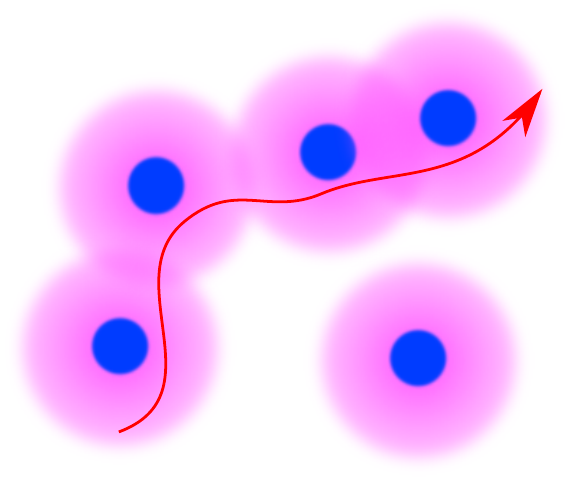}
  \caption{If the interaction clouds are classically percolated, the
    quark mobility seems not restricted and quarks may classically
    flow over connected blocks of meson clouds.}
  \label{fig:dense}
\end{figure}

We point out that Hard Deconfinement is based on hard-core dominance.
The condition for the hard-core dominance is stronger than needed for
deconfinement in a  more conventional sense as conjectured by the
notion of quark mobility. Let us consider decreasing the baryon
density and explore how the quark mobility would change.  It appears
that quarks (accompanied by antiquarks) can still hop from one nucleon
to another through the exchange of mesons.  This situation can be
intuitively understood as overlaps of meson clouds as illustrated in
Fig.~\ref{fig:dense}.  Such an interpretation of the quark mobility
is, however, too na\"{i}ve.  The equilibrium binding of nuclear matter
at the saturation density is sustained mainly by mesonic exchanges,
but needless to say, nuclear matter at the saturation density is not
quark matter yet.  Quark exchanges inevitably occur together with
antiquarks to form color-singlets, and connected blocks of meson
clouds do not really signify quark liberation.

\begin{figure}
  \includegraphics[width=0.8\columnwidth]{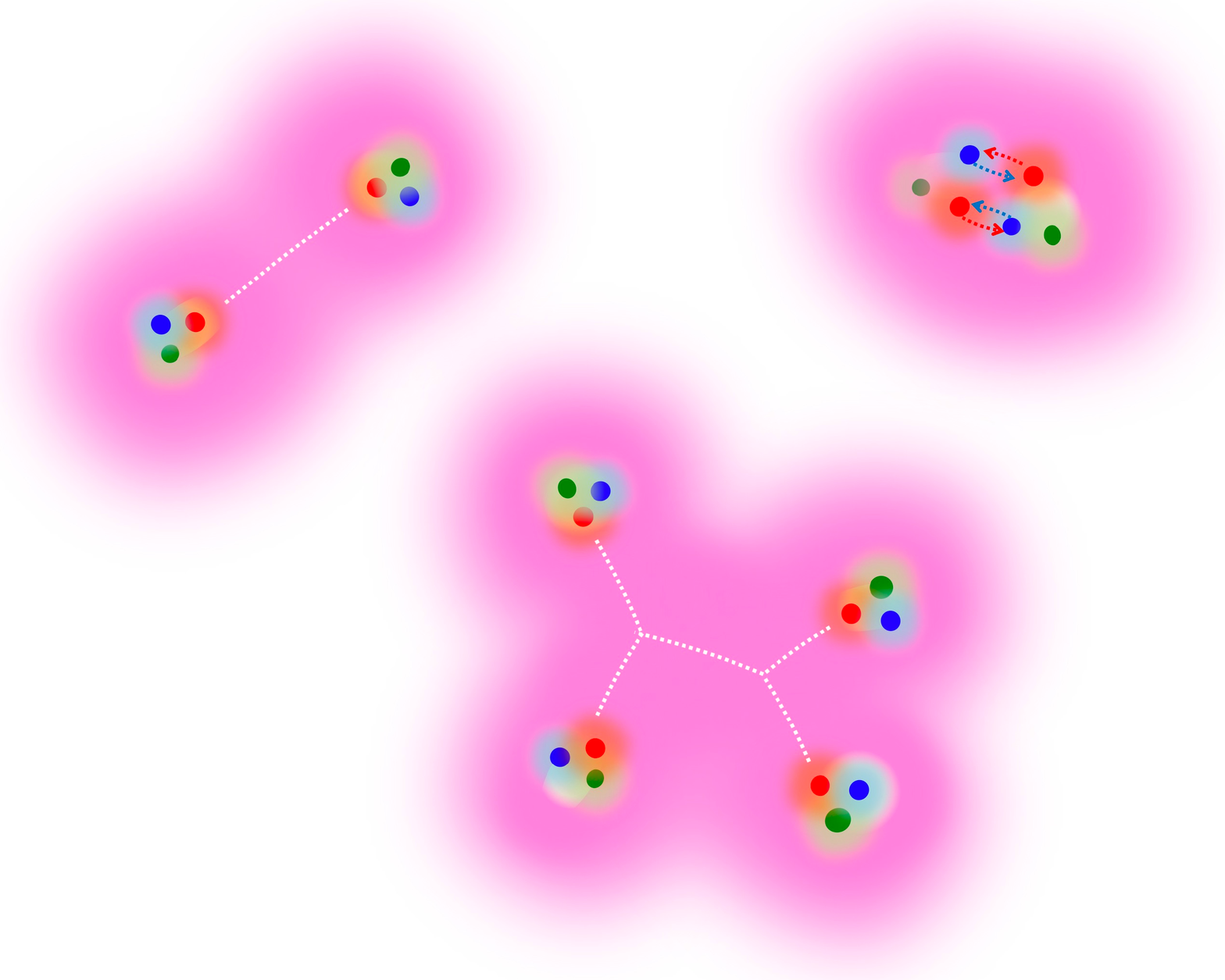}
  \caption{Schematic picture of matter with comparable strengths of  
    all $n$-body interactions.  Extended wave-functions are for quarks
    and antiquarks.  For larger $n$ a picture of individual meson
    exchanges would become more obscured.  At short distances core
    interactions are mediated by quark exchanges.}
  \label{fig:many-body}
\end{figure}

The question that we would like to address in this work is the
following:  there is supposedly another mechanism of quark
deconfinement at lower density than Hard Deconfinement, which we refer
to as \textit{Soft Deconfinement}.  The question is; in other  words,
when does a picture of individual meson exchanges between nucleons
lose its meaning?  If the system is in the confined hadronic phase at
low density, the exchange of color-singlet mesons characterizes baryon
interactions.  The contraposition of this statement is that, if a
meson-exchange based description is blurred, the system should be out
of the confined phase.  Interestingly, this argument suggests a
possible relationship between Soft  Deconfinement and Quarkyonic
Matter.  As mentioned before, Quarkyonic Matter has the potential
energy $\sim \calO(\Nc)$ and all $n$-body interactions are of the same
order.  This is exactly the situation expected in an intermediate
state between nuclear and quark matter in the three-window scenario
description of neutron stars~\cite{Baym:2017whm}.  Even in the real
world with $\Nc=3$ we can still adopt this characterization of
Quarkyonic Matter, namely, matter with comparable strengths of all
$n$-body interactions among nucleons.  From the microscopic point of
view such $n$-body forces could be mediated by multi-meson exchanges
as sketched in Fig.~\ref{fig:many-body}.  In this way we may well
identify Quarkyonic Matter in the $\Nc=3$ real world as multi-body
interacting matter, and we could also adopt this identification for
Soft Deconfinement.

The regime of Soft Deconfinement can thus be viewed as clustering of
nucleons connected by strong $n$-body interactions.  Large $n$ would
imply large clusters.  More precisely, the clusters should be
formulated in terms of wave-functions of quarks and antiquarks.
Mesonic clouds are to be interpreted as ``sea'' quarks which do not
carry net baryon charge.  The corresponding wave-functions of quarks
and antiquarks are equally distributed in space.

Such a spatial extension of wave-functions is quite analogous to those
of electrons in a tight-binding model.  Here, based on an analogy with
condensed matter physics, we are proposing a novel scenario of
deconfinement.  In the metallic state conduction electrons are
extended in space, but a larger concentration of impurity increases
the electric resistivity, and eventually the system under impurity
disturbances behaves as an insulator.  Then, the electron
wave-functions are localized in the insulating state, for which the
physical mechanism is known as Anderson localization.  As a matter of
fact, the idea of the Anderson localization applies to the percolation
problem.  We emphasize that connected blocks of meson clouds in
Fig.~\ref{fig:dense} are percolating classically, but this classical
percolation does not necessarily lead to physical percolation of
wave-functions at the quantum level.

\begin{figure}
  \includegraphics[width=0.45\columnwidth]{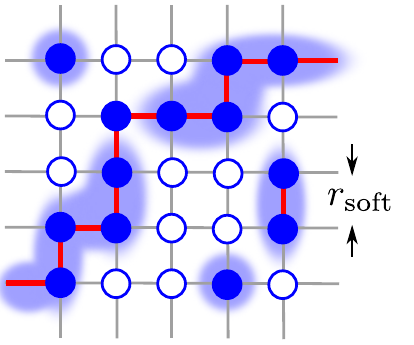}
  \caption{Schematic picture of the classical and the quantum 
    percolation between nucleons.  Neighboring nucleons are linked by 
    interactions shown by red bonds.  The path connected by 
    interaction bonds does not necessarily guarantee extending 
    wave-functions at the quantum level.  The square lattice is only
    for graphical simplicity.}
  \label{fig:grid}
\end{figure}

It is easily understood that the critical concentration for the onset
of percolation should be larger for quantum percolation than for
classical percolation.  The interaction via meson exchanges opens a
classical path for quarks and antiquarks to hop between nucleon sites.
To build a model in the simplest way, let us consider a lattice system
as schematically shown in Fig.~\ref{fig:grid}.  We simplify the
interaction clouds into bonds connecting neighboring sites and place
static nucleons (which is justified in the large-$\Nc$ limit) on
sites.  The bonds should be color-singlets, and this constraint
reduces the strength of the interaction from $\calO(\Nc^2)$ to
$\calO(\Nc)$.  Furthermore, quarks and antiquarks are equally
distributed, reflecting the nature of sea quarks associated with
mesonic clouds.

We increase the baryon density by filling sites randomly, and then,
Soft Deconfinement can be modeled as the site quantum percolation.  In
Fig.~\ref{fig:grid} we see a bond connected path from the left to the
right edge and the classical percolation would allow quarks and
antiquarks to float.  In quantum physics, however, we should solve the
quark and antiquark wave-functions and they could be more localized
than the classical path depending on the corresponding eigenenergy of
the Hamiltonian.  This can be easily understood from familiar examples
of quantum wave-functions that can have nodes.  Such a prominent
distinction between the classical and the quantum percolations is
nicely explained in Ref.~\cite{PhysRevB.45.7724}.  Furthermore, one
essential feature of the quantum percolation is that we should
consider deconfinement of sea quarks and antiquarks mode-by-mode with
different eigenenergies.  Therefore, our proposed quantum percolation
scenario leads to a picture similar to a momentum-shell model of
confinement and deconfinement as proposed in
Refs.~\cite{McLerran:2018hbz,Jeong:2019lhv}.

This paper is organized as follows.  In Sec.~\ref{sec:model} we
discuss the hard core and the soft surface in the nucleon structures.
We introduce a chiral soliton model to demonstrate quantitative
analyses.  As an application we estimate the EoS of dense quark matter
inferred from hard core regions of nucleon in Sec.~\ref{sec:eos}.
Then, we shall proceed to discussions on Soft Deconfinement in
Sec.~\ref{sec:soft}.  In the present paper we limit ourselves to
discuss general properties of quantum percolation and make a
speculative scenario of mode-by-mode Soft Deconfinement.
Section~\ref{sec:conclusions} is devoted to conclusions.

\section{Hard and soft scales in the nucleon}

As a preliminary exercise we shall review the internal structure of
nucleon.  We utilize a chiral soliton model to quantify the hard-core
region where the baryon density is localized, and the soft-tail region
where the meson clouds spread.

\subsection{Empirical facts and phenomenology}
\label{sec:empfacts}

At this point it is useful to give a brief summary of what is known
about scales and sizes of the nucleon, in terms of the empirical radii
determined by various nucleon form factors. Key quantities in this
context are the proton and neutron electromagnetic form factors and
their slopes at zero momentum transfer which define the mean squared
charge radii.  A recent new electron-proton scattering
measurement~\cite{Xiong:2019umf} gives the r.m.s.\ proton charge
radius, $\langle r^2_p\rangle^{1/2} = 0.831\pm 0.014\fm$.  A precision
analysis of deuteron form factors using chiral effective field
theory~\cite{Filin:2019eoe} determines the slope of the neutron
electric form factor as
$\langle r^2_n\rangle = -0.106\pm 0.006\fm^2$.  The isoscalar and
isovector radii of the nucleon, given by
\begin{equation}
  \langle r_{S,V}^2\rangle = \langle r^2_p\rangle \pm \langle r^2_n\rangle \,,
\end{equation}
have the resulting values:
\begin{equation}
  \sqrt{\langle r_{S}^2\rangle} \approx 0.77 \fm\,,
  \qquad
  \sqrt{\langle r_{V}^2\rangle} \approx 0.89 \fm\,.
\label{eq:rSV}
\end{equation}
The isovector charge radius reflects the interacting two-pion cloud of
the nucleon with its spectrum governed by the $\rho$ meson and a
low-mass tail extending down to the $\pi\pi$ threshold.  The isoscalar
charge radius is related to the three-pion
spectrum~\cite{Kaiser:2019irl} that is strongly dominated by the
$\omega$ meson with its mass, $m_V = 783\MeV$, and its narrow width,
$\Gamma = 8.5\MeV$.  The isoscalar electromagnetic current of the
nucleon is well described by the vector meson dominance
phenomenology.  The vector meson dominance relates the electric form
factor, $G_S^E(q^2)$ (with $G_S^E(0)=1$), to a combination of
$F_B(q^2)$ (i.e., the form factor of the baryon number distribution in
the nucleon core) and the $\omega$ field propagation (for which the
baryon density acts as a source):
\begin{equation}
  G_S^E(q^2) = {F_B(q^2)\over 1+|q^2|/m_\omega^2}\,.
\end{equation}
This implies,
\begin{equation}
  \langle r^2_S\rangle = \langle r^2_B\rangle + {6\over m_\omega^2}
\end{equation} 
with the mean squared radius,
\begin{equation}
  \langle r^2_B\rangle = -6{dF_B(q^2)\over dq^2}\Big|_{q^2=0}
\end{equation}
of the baryon density distribution.  With the empirical input one finds,
\begin{equation}
  \sqrt{\langle r^2_B\rangle} \approx 0.45 \fm\,. 
\label{eq:rB}
\end{equation}
The characteristic smallness of the radius of the baryon density
distribution as compared to radius of the charge distribution
underlines the proposed delineation of hard and soft scales in the
nucleon: the valence quarks carrying baryon number are localized in
the compact $\sim 0.5\fm$ core, while quark-antiquark pairs with no
net baryon number form the meson cloud at the nucleon surface.

The sizes and scales just outlined above refer to the vector currents
of quarks in the nucleon.  Another quantity of interest is the scalar
quark density and the corresponding scalar-isoscalar form factor
denoted as $\sigma(q^2)$.  Its value at $q^2 = 0$ is identified with
the pion-nucleon sigma term, $\sigma_{\pi N}$, i.e., the measure of
quark mass contributions to the nucleon mass.  The dispersion relation
representation of $\sigma(q^2)$ involves the two-pion spectral
function related to the $s$-wave isoscalar $\pi\pi$ scattering
amplitude in combination with pion-nucleon scattering.  While details
of a series of investigations of $\sigma(q^2)$ over several
decades~\cite{Gasser:1990ce,Gasser:1990ap,Schweitzer:2003sb,Hoferichter:2015tha}
depend on the precise value of $\sigma_{\pi N}$, there is agreement
that the radius associated with the scalar-isoscalar two-pion cloud of
the nucleon is as large as
$\langle r^2_\sigma\rangle^{1/2} \approx 1.0 - 1.2\fm$. 

The core-plus-cloud picture of the nucleon just discussed and sketched
in Fig.~\ref{fig:dilute} actually arises as a natural consequence of
the spontaneously broken chiral symmetry of QCD at low
energies~\cite{Brown:1985gu}.  The compact hard core hosting the
valence quarks, and the surrounding soft surface composed of
Nambu-Goldstone bosons (the multi-pion cloud), are the basic
components of a variety of chiral models of the
nucleon~\cite{Thomas:2001kw}, ranging from different versions of
chiral and cloudy bag models to chiral solitons.  We will use the
latter for orientation in order to quantify the hard and soft scales
in the nucleon.

\subsection{Hard core and soft surface in a chiral soliton model}
\label{sec:model}

\begin{figure}
  \includegraphics[width=0.97\columnwidth]{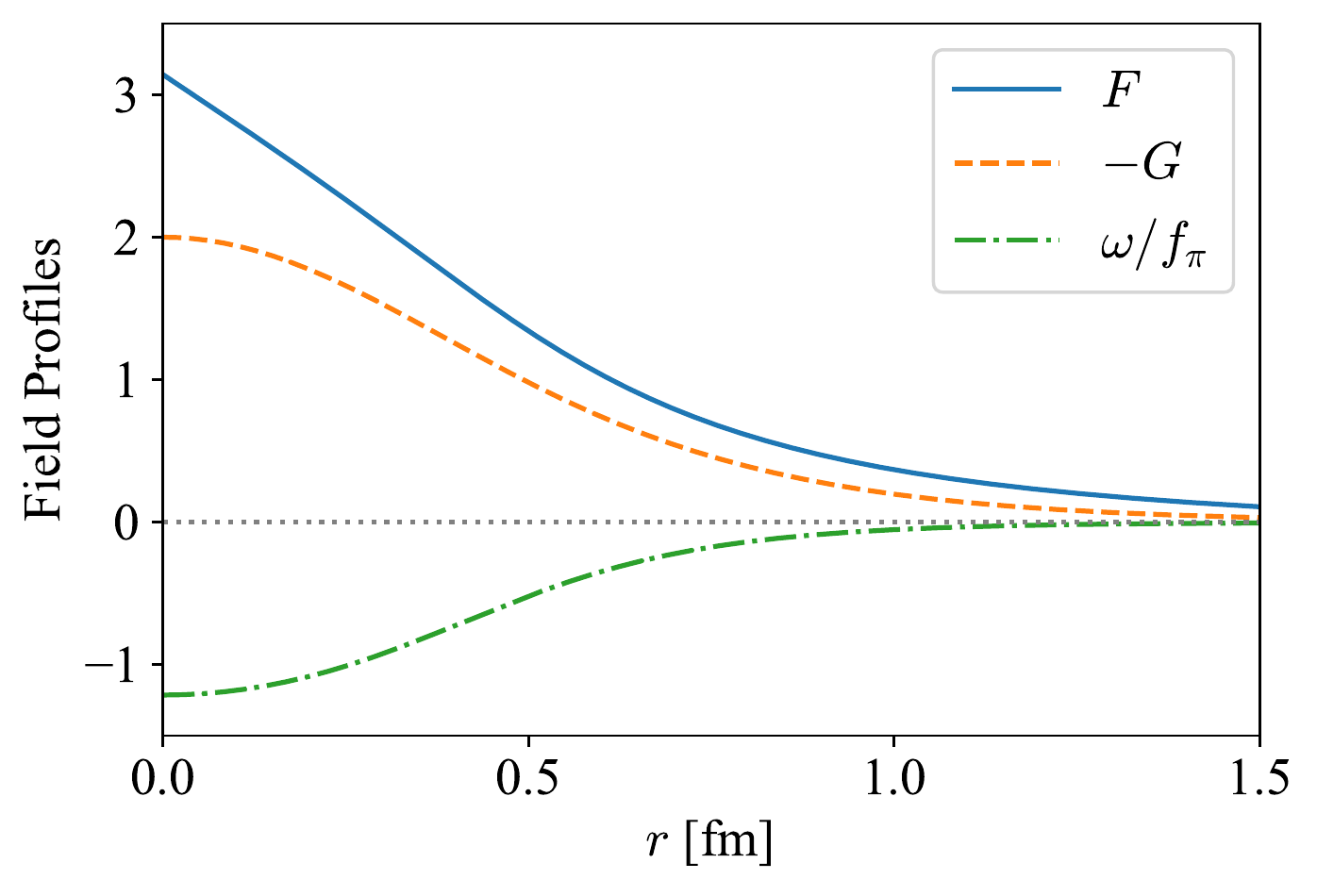}
  \caption{Numerical solutions of $F(r)$ (solid curve), $G(r)$ (dashed 
    curve), and $\omega(r)$ (dot-dashed curve) scaled dimensionlessly
    with $f_\pi$.}
    \label{fig:FGw}
\end{figure}

We shall now look into the baryon structure more quantitatively.  To
this end we choose a chiral soliton model with $\pi$, $\rho$, and
$\omega$ fields.  For simplicity we will not quantize the soliton but
rescale the results by a mass discrepancy as we will explain later.
According to Refs.~\cite{Meissner:1986ka,Meissner:1986js}, we can
construct baryons as the Skyrmions of $\pi$ field which is stabilized
by not the Skyrme term but the coupling with $\rho$ and $\omega$
vector mesons as first considered in Ref.~\cite{Adkins:1983nw}
followed shortly by Ref.~\cite{Igarashi:1985et}.  We introduce the
chiral fields and the vector fields as
\begin{align}
  U(\br) &= e^{i\btau\cdot \hat{\br}\,F(r)} \,,\\
  \rho^{i,a}(\br) &= \epsilon^{ika} \hat{\br}^k \frac{G(r)}{gr}\,,\\
  \omega^\mu(\br) &= \delta^{\mu 0}\omega(r) \,.
\end{align}
We determine these fields to minimize the energy, i.e., these fields
should satisfy a set of equations as 
\begin{align}
  F'' &= -\frac{2}{r}F' + \frac{1}{r^2} \bigl[ 4(G+1)\sin F - \sin 2F
        \bigr] \notag\\
      &\qquad\qquad\qquad
        + \tilde{m}_\pi^2 \sin F
        - \frac{3g}{4\pi^2 r^2} \omega' \sin^2 F\,,\\
  G'' &= 2g^2 (G+1-\cos F) + \frac{1}{r^2}G(G+1)(G+2)\,,\\
  \omega'' &= -\frac{2}{r}\omega' + 2g^2 \omega
             - \frac{3g}{4\pi^2 r^2} F' \sin^2 F\,.
\end{align}
In what follows below we combine $r$, $\omega$, and $m_\pi$ with
$f_\pi$ to make them dimensionless (where the dimensionless pion mass
is specifically denoted by $\tilde{m}_\pi$).  In the above equations
only one mass scale independent of $f_\pi$ is the pion mass $m_\pi$,
while the vector meson masses,
$m_\rho^2=m_\omega^2=2g^2 f_\pi^2$, follow from the KSFR
(Kawarabayashi-Suzuki~\cite{PhysRevLett.16.384.3} and
Fayyazuddin-Riazuddin~\cite{PhysRev.147.1071}) relation and they scale
with $f_\pi$.  From the KSFR relation we can fix
$g=m_\omega/(\sqrt{2}f_\pi)=6.0$ (with
$\sqrt{2}f_\pi = 130.41\MeV$ and $m_\omega = 783\MeV$).  We solve
these differential equations under the boundary conditions:
\begin{equation}
  F(0) = \pi\,,\qquad F(\infty) = 0\,,
\end{equation}
which is required to quantize the baryon number $B=1$.  The vector
mesons should satisfy the following boundary conditions:
\begin{equation}
  \begin{split}
  & G(0) = -2\,,\qquad G(\infty) = 0\,,\\
  & \omega'(0) = 0\,,\qquad \omega(\infty) = 0\,.
  \end{split}
\end{equation}
We note that $F(0)$, $G(0)$, and $\omega'(0)$ are initial conditions
and we adjust the rest of initial conditions, $F'(0)$, $G'(0)$, and
$\omega(0)$, to realize the asymptotic behavior,
$F(\infty), G(\infty), \omega(\infty)\to 0$.  We show the numerical
solutions for $F(r)$, $-G(r)>0$, and (dimensionlessly scaled)
$\omega/f_\pi$ in Fig.~\ref{fig:FGw}.

With these field profiles we can immediately compute physical
quantities.  In this way we can concretely demonstrate the baryon
structure and exemplify a hard core surrounded by a soft tail.  The
baryon number density $\rhoB(r)$ should be localized in the hard core
region and we can see this from an explicit expression:
\begin{equation}
  \rhoB(r) = -\frac{1}{2\pi^2 r^2} F'(r) \sin^2 F(r) \,,
\end{equation}
which leads to the properly quantized baryon number;
$B=4\pi \int dr\, r^2\rhoB(r)=1$.  The pion clouds can be
characterized by the isoscalar charge density given by
\begin{equation}
  \rhoS(r) = -2 g f_\pi^2 \omega\,.
\end{equation}

\begin{figure}
  \includegraphics[width=\columnwidth]{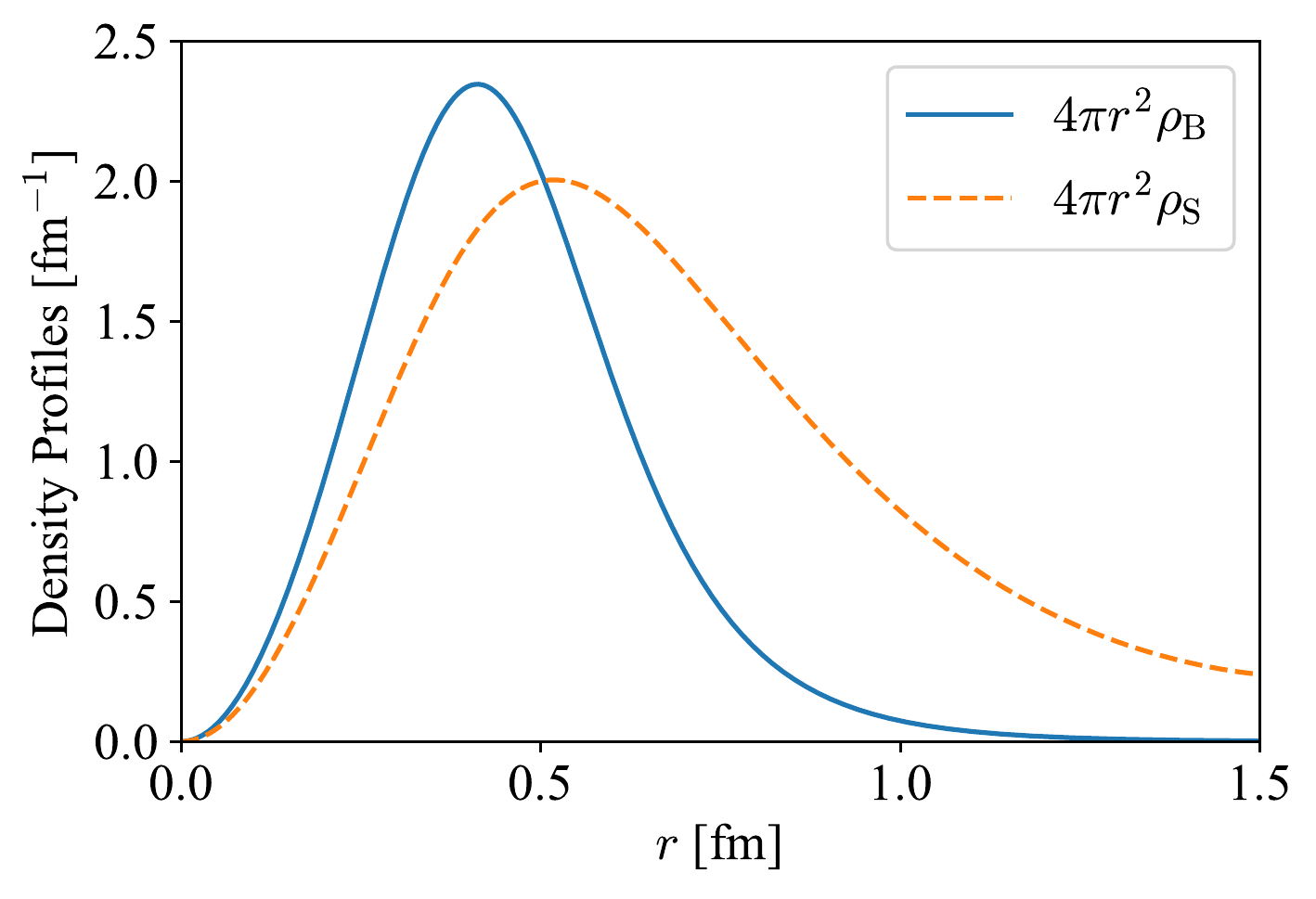}
  \caption{Baryon and isoscalar charge density
    distributions as functions of $r$ multiplied by $4\pi r^2$.}
    \label{fig:density_rsq}
\end{figure}

We show the numerical behavior of $4\pi r^2\rhoB(r)$ and
$4\pi r^2\rhoS(r)$ in Fig.~\ref{fig:density_rsq}, where $\rhoB(r)$
gives the information on the core extension and $\rhoS(r)$ has a
longer tail which behaves asymptotically as $\sim e^{-3m_\pi r}$.
Using these distributions we can estimate the r.m.s.\ radii as
\begin{align}
  \sqrt{\langle r^2_B\rangle}
  &= \left(   \frac{\int_0^\infty dr\, r^4\, \rhoB(r)}
    {\int_0^\infty dr\, r^2\, \rhoB(r)}  \right)^{1/2} \approx 0.49\fm\,, \\
  \sqrt{\langle r^2_S\rangle}
  &= \left(  \frac{\int_0^\infty dr\, r^4\, \rhoS(r)}
    {\int_0^\infty dr\, r^2\, \rhoS(r)}\right)^{1/2}  \approx 1.03\fm\,.
    \label{eq:r2s}                              
\end{align}
These numbers are close to those of Eqs.~\eqref{eq:rSV} and
\eqref{eq:rB}.  It is conceivable to relate
$\sqrt{\langle r^2_B\rangle}$ to $\rhard$ and
$\sqrt{\langle r^2_S\rangle}$ to $\rsoft$.

It should be noted here that $\rhoB(r)$ represents the net baryon
distribution and thus the hard core should be dominated by ``valence''
quark degrees of freedom, as already mentioned.  In contrast, the
interaction clouds carry no net baryon charge, and the interaction
tail should be dominated by ``sea'' quark degrees of freedom.  This
difference is a key ingredient to distinguish two states of
deconfinement in later discussions.

\section{Hard deconfinement and the equation of state}
\label{sec:eos}

Here we discuss the nature of Hard Deconfinement and its implication
to the EoS of dense quark matter.  As we already mentioned, the quark
mobility itself could be enhanced even before the hard cores touch
each other, and strictly speaking, there is no transition of
deconfinement associated with Hard Deconfinement.  Still, it is
convenient to think of matter in the regime of Hard Deconfinement.

The term Hard Deconfinement is used to consider a state of matter
dominated by properties of the baryon hard cores.  Let us first give a
rough estimate of the relevant density for Hard Deconfinement.  We
shall assume the closest packed state, i.e., either the hexagonal
close-packed (hcp) or the face-centered cubic (fcc) lattice, in which
the filling rate is $74\%$.  As we read from
Fig.~\ref{fig:density_rsq} the hard core radius is around
$\rhard \sim 0.5\fm$.  When the closest packed state occurs, the
baryon density corresponding to the filling fraction is
\begin{equation}
  \sim \;\; 0.74 \; \times\ \biggl(\frac{4}{3}\pi \rhard^3\biggr)^{-1}
  \;\; \simeq \;\; 1.4\fm^{-3}\,,
\label{eq:filling}
\end{equation}
which is $\sim 8.3\rho_0$ in the unit of the normal nuclear density;
$\rho_0\sim 0.17\fm^{-3}$.  We note that this density,
$\sim 8.3\rho_0$, is estimated for the closest packed state, and so
this should be taken as the limiting value below which Hard
Deconfinement should be realized.  It is a sufficient condition that
the hard cores (as rigid spheres) touch each other in the static
closest packed state. In reality nucleons have Fermi motion, and their
hard cores are to be replaced  by continuous distributions as seen in
Fig.~\ref{fig:density_rsq}.  The critical density for Hard
Deconfinement as a continuous transition can thus be lower than
Eq.~\eqref{eq:filling} by some factor.

\begin{figure}
  \includegraphics[width=\columnwidth]{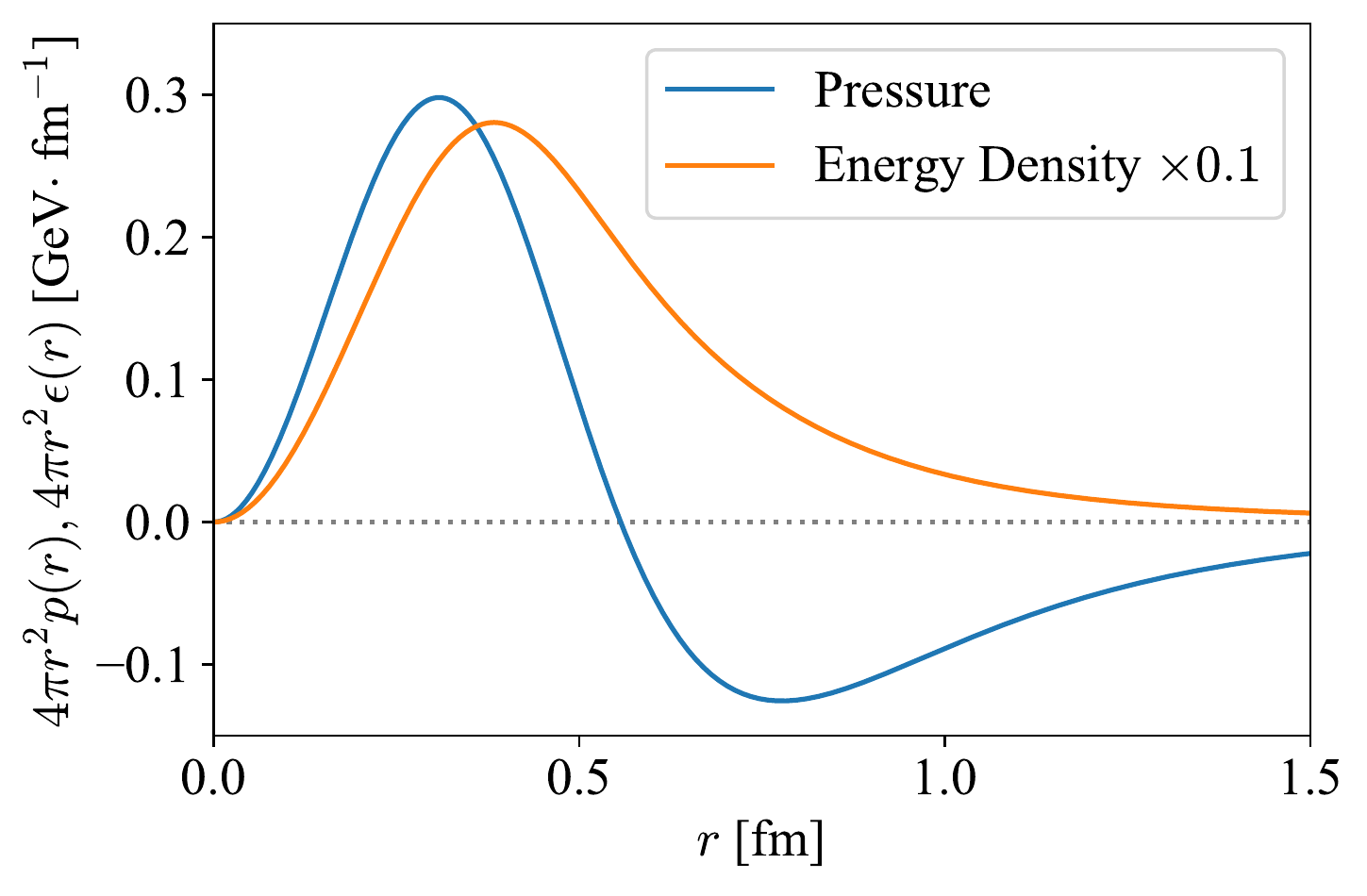}
  \caption{Pressure and energy density distributions as functions of 
    $r$ multiplied by $4\pi r^2$.  To make the comparison easier, the 
    energy density is rescaled by a factor $0.1$.}
    \label{fig:PE_sq}
\end{figure}

Hard Deconfinement provides us with an interesting opportunity to make
a quantitative assessment of bulk matter from a single baryon.  
It is conceivable for thermodynamic pressure of bulk matter 
to be approximated by 
 {\it mechanical} pressure in a hard core,
%
$ p (x) = \langle N| T_{ii} (x) | N \rangle$, 
where $|N\rangle$ is a nucleon state and $T_{\mu \nu}$
is the energy momentum tensor. 
To separate the surface effects specific to an isolated nucleon,
we focus on the pressure near the center of the hard core.
Within the framework of the model
explained in Sec.~\ref{sec:model}, we can compute the pressure and the
energy density.  
Their expressions (made dimensionless divided by
$f_\pi^4$) are found to be
\begin{widetext}
  \begin{align}
    p(r) &= -\frac{1}{6}F^{\prime 2} - \frac{\sin^2 F}{3r^2}
    - \frac{2}{3}\frac{(G+1-\cos F)^2}{r^2} + g^2 \omega^2
    + \frac{1}{6}\omega^{\prime 2}
    + \frac{G^{\prime 2}}{3g^2 r^2} + \frac{G^2(G+2)^2}{6g^2 r^4}
    - \tilde{m}_\pi^2 (1-\cos F) \,,\\
    \epsilon(r) &= \frac{1}{2} F^{\prime 2} + \frac{\sin^2 F}{r^2}
    +2\frac{(G+1-\cos F)^2}{r^2} - g^2\omega^2
    - \frac{1}{2}\omega^{\prime 2} + \frac{G^{\prime 2}}{g^2 r^2}
    + \frac{G^2(G+2)^2}{2g^2 r^4}
    + \frac{3g}{4\pi^2 r^2} \omega F' \sin^2 F
    + \tilde{m}_\pi^2 (1-\cos F) \,.
  \end{align}
\end{widetext}
We can check from the Virial theorem that the above pressure
satisfies; $\int dr\, r^3 p(r) = 0$ as it should.  We performed
numerical calculations and present $p(r)$ and $\epsilon(r)$
(multiplied by $4\pi r^2$) in Fig.~\ref{fig:PE_sq} where we rescaled
the energy density by a factor $0.1$ to make it comparable to the
pressure.  The characteristic feature of the pressure distribution
inside the nucleon is its combination of a positive core pressure and
a negative pressure at the surface~\cite{Cebulla:2007ei,Goeke:2007fp},
adding up to overall zero pressure to maintain equilibrium in the
nucleon ground state.  Such a pressure profile is verified, at least
qualitatively, in deeply virtual Compton scattering
measurements~\cite{Burkert:2018bqq}.

From these results one can infer 
the relation between the mechanical pressure and energy density
in the core region of the
nucleon, which may serve as a reasonable approximation for the EoS of
quark matter near the closest packed density~\eqref{eq:filling}.  One
might care about differences between symmetric nuclear matter and
neutron matter, but in such an extremely high-density regime of our
interest the physical properties are to be dominated by the strong
interaction and the $\beta$-equilibrium condition would be not
essential.

In the present framework we must be careful of the mass scale in
executing this program for the EoS construction.  As discussed in the
previous work~\cite{Meissner:1986ka,Meissner:1986js}, this chiral
soliton model overestimates the baryon mass which is given by the
integration of the energy density.  It is known that this mass
discrepancy would be reduced if the soliton is quantized (i.e.,
rotated with spin and isospin).  Here, our main purpose is not to
study the chiral soliton model itself but to demonstrate the idea, so
we shall adopt a quick prescription:  we rescale the results simply by
the ratio, $\chi$, between the physical nucleon mass and the model
output.   That is, we introduce a ratio parameter as
\begin{equation}
  \chi = \frac{\text{(physical mass)}}{\text{(model mass)}}
  \approx \frac{940\MeV}{1460\MeV} \approx 0.64\,.
\end{equation}
Then, we should make the following rescaling:
\begin{equation}
  \epsilon(r)\;\to\; \chi \epsilon(r)\,,\qquad
  p(r)\;\to\; \chi^{-1} p(r)\,.
\end{equation}
The above is the consistent rescaling in such a way not to modify the
form factors.  In other words, given the nucleon form factors $A(q^2)$
and $D(q^2)$ associated with the components of the energy-momentum
tensor, the energy density is proportional to the mass, while the
pressure is inversely proportional to the mass (see a
review~\cite{Polyakov:2018zvc} for explicit expressions).  One might
have thought that the model parameters can be readjusted to fit the
baryon mass, but this would significantly affect the charge radius.
If the form factors stay intact leaving the charge radius unchanged,
the rescaling procedure should yield physically more sensible results
than readjusting the model parameters.

\begin{figure}
  \includegraphics[width=\columnwidth]{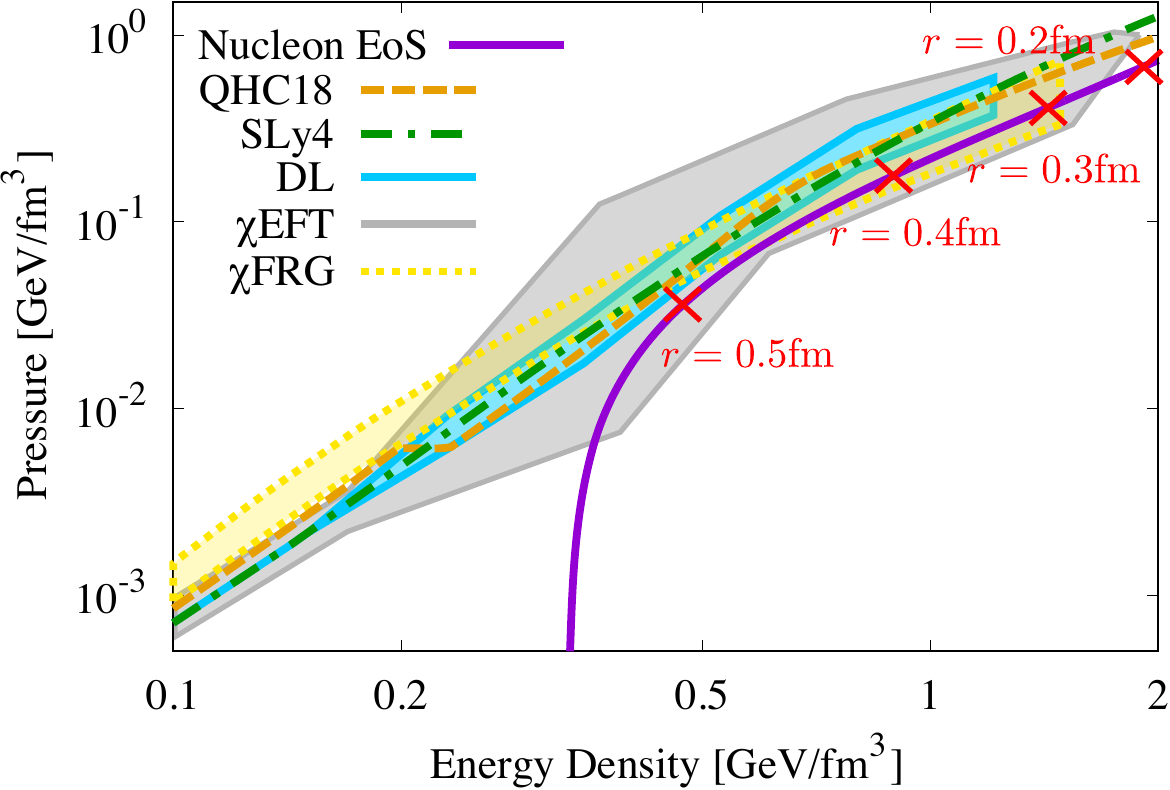}
  \caption{EoS of dense quark matter from the Hard Deconfinement 
    scenario (Nucleon EoS) and empirical EoSs from other approaches.}
    \label{fig:EoS}
\end{figure}

Figure~\ref{fig:EoS} presents our results for the equation of state,
$p(\epsilon)$, of dense quark matter in the hard core region of the
nucleon, compared to several proposed EoSs that are consistent with
empirical properties of neutron star matter.  We label our results,
the rescaled $p(r)$ and $\epsilon(r)$, as ``Nucleon EoS'' and mark
different radial coordinate scales in the nucleon core, $r=0.2\fm$ to
$0.5\fm$, with crosses in Fig.~\ref{fig:EoS}.  The fast-dropping
behavior at $r \gtrsim 0.5\fm$ reflects the negative pressure at the
nucleon surface, physically interpreted as resulting from confining
forces and the inward-bound pressure of the meson cloud.

For the neutron star based equations-of-state in Fig.~\ref{fig:EoS},
$\chi$EFT refers to the EoS from the Chiral Effective
Theory~\cite{Hebeler:2013nza} and QHC18 from Ref.~\cite{Baym:2017whm},
and SLy4 from Ref.~\cite{Douchin:2001sv}.  DL shows the EoS deduced
from the observation data analyses using the deep
learning~\cite{Fujimoto:2019hxv}.  The EoS data labelled by $\chi$FRG
is taken from Refs.~\cite{Drews:2014spa,Friman:2019ncm}.  We note that
the EoS bound from the deeply virtual Compton scattering on the proton
was previously discussed in a similar way in
Ref.~\cite{Liuti:2018ccr}.  For $r < 0.5 \fm$, remarkable agreement is
seen between our (free) Nucleon EoS and the sets of dense neutron star
matter equations-of-state.  Assuming that the onset of Hard
Deconfinement appears at $r$-scales in the range $r = 0.5 - 0.4 \fm$
(corresponding to baryon densities $\sim 4 - 7 \rho_0$ according to
Fig.~\ref{fig:density_rsq}), this implies that Hard Deconfinement can
occur at significantly lower density than the limiting
estimate~\eqref{eq:filling}.

Before closing this section we mention the possibility of partial
chiral symmetry restoration at high density.  So far we have been
looking at single baryon properties in vacuum, but interactions with
surrounding baryons are expected to change its properties in a
high-density environment.  The simplest way to implement this effect
is to reduce the chiral order parameter, $f_\pi$, in the chiral
soliton model.  Although the model has two independent mass scales,
$f_\pi$ and $m_\pi$, we numerically found that the solutions of the
chiral soliton model scale with $f_\pi$ in a good approximation.  This
means that, if $f_\pi$  decreases to $f_\pi^\ast$ in a medium,
$\epsilon(r)$ and $p(r)$ change as
\begin{equation}
  \epsilon(r) \;\to\;  \biggl(\frac{f_\pi^\ast}{f_\pi}\biggr)^4\, \epsilon(r)\,,
  \qquad
  p(r) \;\to\; \biggl(\frac{f_\pi^\ast}{f_\pi}\biggr)^2\, p(r)\,.
\end{equation}
In Fig.~\ref{fig:EoS} these modifications shift $\epsilon(r)$ by a
factor along the horizontal axis, while the vertical axis is given by
the logarithmic scale and the vertical shift of the Nucleon EoS curve
is not by a factor but nearly by an offset.  Therefore, the EoS would
be stiffer if in-medium $f_\pi^\ast$ gets smaller with increasing density.  
This stiffening, with $f_\pi^\ast\sim 0.8f_\pi$ for example,
improves the agreement between ``Nucleon EoS'' and the others in the
high density region in Fig.~8.
We note that the ratio $\chi$ might in principle also be
density dependent, but this dependence should be approximately
negligible.  This is because both the physical mass and the model mass
should be affected by partial chiral symmetry restoration and their
ratio is expected to be unchanged.

\section{Soft deconfinement as quantum percolation}
\label{sec:soft}

In this section we discuss the scenario of Soft Deconfinement.  We
begin with a brief description of classical percolation and then
proceed to a concrete model of quantum percolation.  We summarize the
basic properties of quantum percolation and translate them in the
context of quark liberation in dense baryonic matter.  Most important
is the observation of energy dependent percolation which leads to a
novel picture of mode-by-mode delocalization of quark wave-functions,
akin to a recently proposed momentum-shell model in the Quarkyonic
Matter picture.

\subsection{Classical percolation}

As we did previously for Hard Deconfinement, we shall begin with an
order estimate for the critical concentration in percolating baryonic
matter.  The idea to interpret quark deconfinement as percolating
baryons is traced back to pioneering works, see
Ref.~\cite{Karsch:1979zt} for example, in which the ``hadron solid''
could be considered in terms of weakly coupled quark matter.  This is
a prototype of the percolation model of quark deconfinement.  A more
refined picture was discussed in Ref.~\cite{Castorina:2008vu}, in
which the percolating density was estimated as $\sim 5.5\rho_0$.  
The $\Nc$ dependence of percolation for tightly packed baryons was discussed
in Ref.~\cite{Lottini:2011zp}.  
In our language these percolation models with hard cores rather
correspond to Hard Deconfinement.

The characteristic scale in the Soft Deconfinement scenario is given
by a typical length scale of the mesonic clouds rather than the hard
cores.  Such a picture was also considered, for example, in
Ref.~\cite{Baym:2008me}.  The essential argument in
Ref.~\cite{Baym:2008me} is that the relative importance of multi-body
interactions is given parametrically by $n/(2m_\pi)^3$ where $n$
represents the baryon density, and this approaches $\sim\calO(1)$ when
$n$ gets larger, which is also emphasized in the recent
review~\cite{Baym:2017whm}.  We note, however, that the relevant
length scale is not necessarily $\sim 1/(2m_\pi)$ in reality, and the
multi-body interactions in the $\chi$EFT suggest that the relevant
scale should be $1.1\sim 1.3\fm$~\cite{Kaiser:2012ex}, which is
consistent with the r.m.s.\ radius of the scalar-isoscalar nucleon
form factor~\cite{Gasser:1990ap} (depending on the value of the
pion-nucleon sigma term).  Also, if the relevant scale is related to
the Compton wavelength corresponding to a spectral maximum in the
scalar-isoscalar channel, it would be $1/m_\sigma\sim 0.4\fm$.  The
soft scale has the largest uncertainty in this picture and in the
present consideration we shall choose $\rsoft \sim 0.7\fm$ for the
moment, close to the value in Eq.~\eqref{eq:rSV}.

Now, let us concretize the percolation picture by the following
modeling.  We assume the Born-Oppenheimer approximation; (i) Baryons
move at velocity $p_F/m_B \sim \calO(\Nc^{-1})$, much slower than
quarks at velocity of $\calO(1)$.  (ii) Quark wave-functions solved
for a given quasi-static configuration of baryons.  (iii) Physical
quantities estimated as a result of averaging over the slow baryon
dynamics.  For simplicity we replace the time averaging procedure in
(iii) by the ensemble average.  Within this framework we shall give a
quick estimate of the critical percolating density adopting a
three-dimensional model of sphere percolation.  In this model there
are sites connected by networks (i.e., connected bonds), and each site
is either occupied by a particle (baryon) with the probability $p$, or
empty with the probability $1-p$.  
We note that the model allows one baryon per site
and this feature is consistent with the hard core repulsion.
The probability $p$ can be easily
translated into the particle density.

The classical percolation in this three-dimensional model is
characterized by the existence of connecting networks between
different boundaries of the whole system. 
The critical filling fraction $p_{\rm c}$ is defined by a condition
that a cluster connecting two boundaries (e.g., $x=-\infty$ and
$x=+\infty$) begins to be formed.
%
 It is known from
Ref.~\cite{doi:10.1063/1.4742750,*doi:10.1063/1.4898557} that the
critical filling fraction is $\pc \sim 0.34$ in this model (other
types of lattice models take a similar value of the critical filling
fraction).  Thus, the critical density for classical percolation would
be
\begin{equation}
  \sim \;\; 0.34 \;\;\times\;\; \biggl(\frac{4}{3}\pi \rsoft^3\biggr)^{-1}
  \;\;\simeq\;\; 0.24 \fm^{-3}\,,
  \label{eq:fillingsoft}
\end{equation}
assuming $\rsoft\sim 0.7\fm$.  We emphasize that this estimate is just
for qualitative considerations:  if $\rsoft$ were slightly changed by
hand, the above number would quantitatively differ.  With this caution
in mind, using Eq.\,\eqref{eq:fillingsoft} the critical density for
percolation would be $\sim 1.4\rho_0$.  It is obviously unlikely to
expect quark matter to appear at such low density.

We note that the meson clouds saturate the system at $p=1$ which corresponds to $\rho \sim 4.2 \rho_0$.
At this density baryons may be still non-relativistic, provided that meson clouds do not strongly limit the motion of hard 
cores\footnote{ 
For the combined effects of scalar-isoscalar ($\sigma$) boson and $\omega$ meson clouds, each
contribution is $\calO(\Nc)$, but they largely cancel.
Hence we assume that short-range repulsion sets in at the hard scale $\sim r_{\rm hard}$.
}.
Toward Hard Deconfinement at $\rho \sim 8.3 \rho_0$, hard core repulsions exclude volumes available for a baryon and lifts up the momenta, making the baryon relativistic.
This is beyond the Born-Oppenheimer descriptions.
The excluded volume thermodynamics for baryons was discussed in Ref.~\cite{Jeong:2019lhv}.

\subsection{Quantum percolation}

Quantum mechanically, the availability of classical quark paths would
not guarantee 
percolated wave-functions, i.e., wave-functions delocalized in one
arbitrary direction connecting two boundaries.
%
This is because there may be
destructive interferences and nodes appearing in quantum amplitudes.
To consider such quantum effects, let us briefly review a simple
quantum percolation model (see Ref.~\cite{Chakrabarti:1339349} for a
comprehensive book; we refer to results in
Ref.~\cite{PhysRevB.6.3598}).  A typical toy model is defined by the
following Hamiltonian of the tight binding model:
\begin{equation}
  H = \sum_n |n\rangle \varepsilon_n \langle n|
  + \sum_{n\neq m} |n\rangle V_{nm} \langle m| \,,
\end{equation}
where $|n\rangle $ denotes a state with a quark occupying site $n$.
The term involving $V_{nm}$ describes quark hopping between sites $n$
and $m$.  For the site percolation problem the simplest choice would
be $V_{nm}= - V~ (V>0)$ for the nearest neighbor sites and $V_{nm}=0$
otherwise.  The background baryon distribution is treated classically
and specified by $\varepsilon_n$.  Each site $n$ is occupied by a
baryon with the probability $p$ or left empty with the probability
$1-p$.  The site energies of quarks, $\varepsilon_n$, depend on
whether the site $n$ is occupied by the baryon or not.  Let us
introduce a notation $\varepsilon_{\rm on}$ to represent the site
energy of quarks at occupied site, that is, a quark energy as a part
of placed baryon.  If the site is left empty, the site energy is
$\varepsilon_{\rm off}$.  In this way, $\varepsilon_n$ is generated by
the following probability distribution:
\begin{equation}
  P(\varepsilon_n) = p \delta(\varepsilon_n -\varepsilon_{\rm on} )
  + (1-p) \delta(\varepsilon_n - \varepsilon_{\rm off}) \,.
\end{equation}
We set $\varepsilon_{\rm off}\rightarrow \infty$ with which quarks
cannot penetrate into empty sites.  We note that this limit is common
in quantum percolation problems (for large but finite
$\varepsilon_{\rm  off}$, see, e.g.,
Refs.~\cite{PhysRevB.6.3598,PhysRevB.45.7724}).  Under this limit the
classical percolation should be a necessary condition for the quantum
percolation; the critical concentration for quantum percolation,
$\pc$, must satisfy $\pq \ge \pc$.  The eigenenergy $E$ of the
Hamiltonian for this single particle problem is to be regarded as the
kinetic energy of a (non-relativistic) quark.  The kinetic energy thus
depends on the baryon cluster size.  Let us consider two extreme
examples.  For a quark localized in a single baryon the eigenenergy is
$\varepsilon_{\rm on}$ and from the uncertainty relation this is the
high energetic case.  Conversely, for completely percolated baryons
with $p=1$, the eigenstates are plane-waves and the eigenenergies are
$E(k) = \varepsilon_{\rm on}  - 2V \sum_{i=x,y,z} \cos(k_i)$.
Here, let us choose the energy offset so that
$\varepsilon_{\rm on} = 6V$ and
$E(k) = 4V  \sum_{i=x,y,z} \sin^2 (k_i/2) $.  This choice is
physically reasonable for our purpose and the kinetic energy vanishes
as $k_{i} \rightarrow 0$, while $\varepsilon_{\rm on}=0$ is a usual
choice in condensed matter literatures on quantum percolation.

\begin{figure}
  \includegraphics[width=\columnwidth]{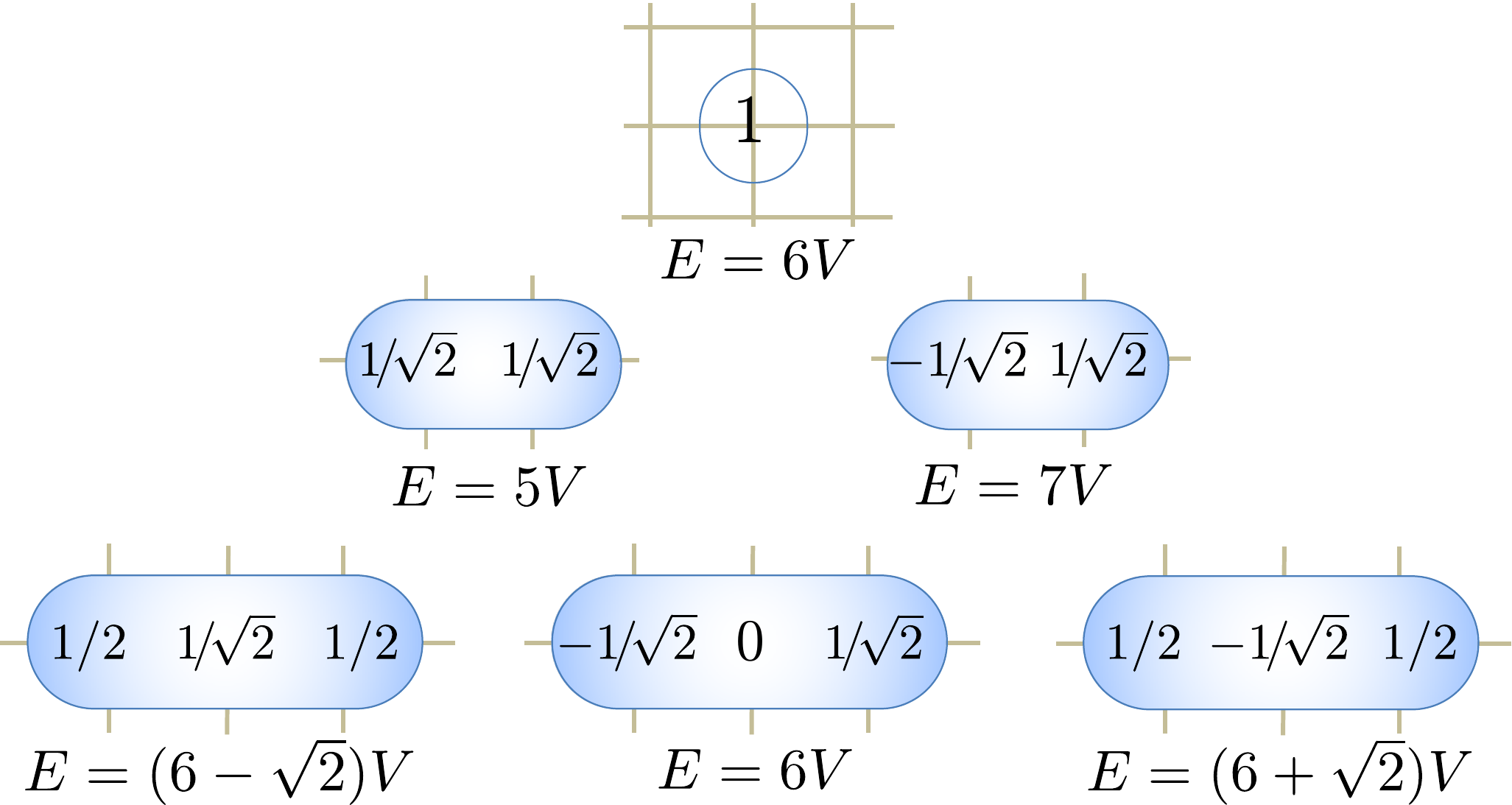}
  \caption{Examples of single particle states for one-dimensionally
    aligned baryon configurations.  The numbers attached to sites are the
  quantum amplitudes.  The quark eigenenergy on an isolated baryon is
  $\varepsilon_{\rm on}=6V$.  With two-baryon connected cluster (as
  shown in the middle row) a more extended wave-function is allowed
  and the lowest quark eigenenergy is lowered.  The three-baryon
  cluster (as shown in the bottom) has a state of eigenenergy $6V$ for
  which a node separates the cluster into two ``localized''
  wave-functions.}
    \label{fig:examples_clusters}
\end{figure}

Since the quantum interference is sensitive to the wavelength, $\pq$
is a function of $E/V$ and further depends on geometrical site
structures (i.e., square lattice, triangular lattice, continuum
spheres, etc).  In Fig.~\ref{fig:examples_clusters}, we illustrate
simple examples to exemplify a ``localized'' wave-function.  For a
clear demonstration purpose let us consider one-dimensionally aligned
baryons from one to three.  The numbers attached to each site are the
amplitudes of the quark wave-functions.  When the amplitudes of two
neighboring domains have opposite signs, there must be a node between
them.  For example, with the two baryon background (as shown in the
middle of Fig.~\ref{fig:examples_clusters}) the quark wave-functions
have zero and one node, respectively, with the eigenenergies $5V$ and
$7V$.  Then, we would call such a state with $E=7V$, which is
partitioned into two, the ``localized'' state.  More precisely
speaking, we define localized states as a finite-amplitude domain
surrounded by vanishing amplitudes (or exponentially suppressed
amplitudes for more general continuum models).  If the whole system is
just three sites and all three sites are occupied by baryons, the
quark state as shown in the left bottom in
Fig.~\ref{fig:examples_clusters} is delocalized over the whole system
and this is our definition of Soft Deconfinement.  In the bottom of
Fig.~\ref{fig:examples_clusters} the central figure shows a
``localized'' state with one node.  Interestingly, this state has
$E=\varepsilon_{\rm on}=6V$, the same energy as the isolated single
baryon case (as shown in the top of Fig.~\ref{fig:examples_clusters}).
The far-right figure is the most energetic state with two nodes.
Again, we emphasize that the quantum interference of reflected waves
with boundaries is essential to create the zeros in the
wave-functions.

\begin{figure}
  \includegraphics[width=\columnwidth]{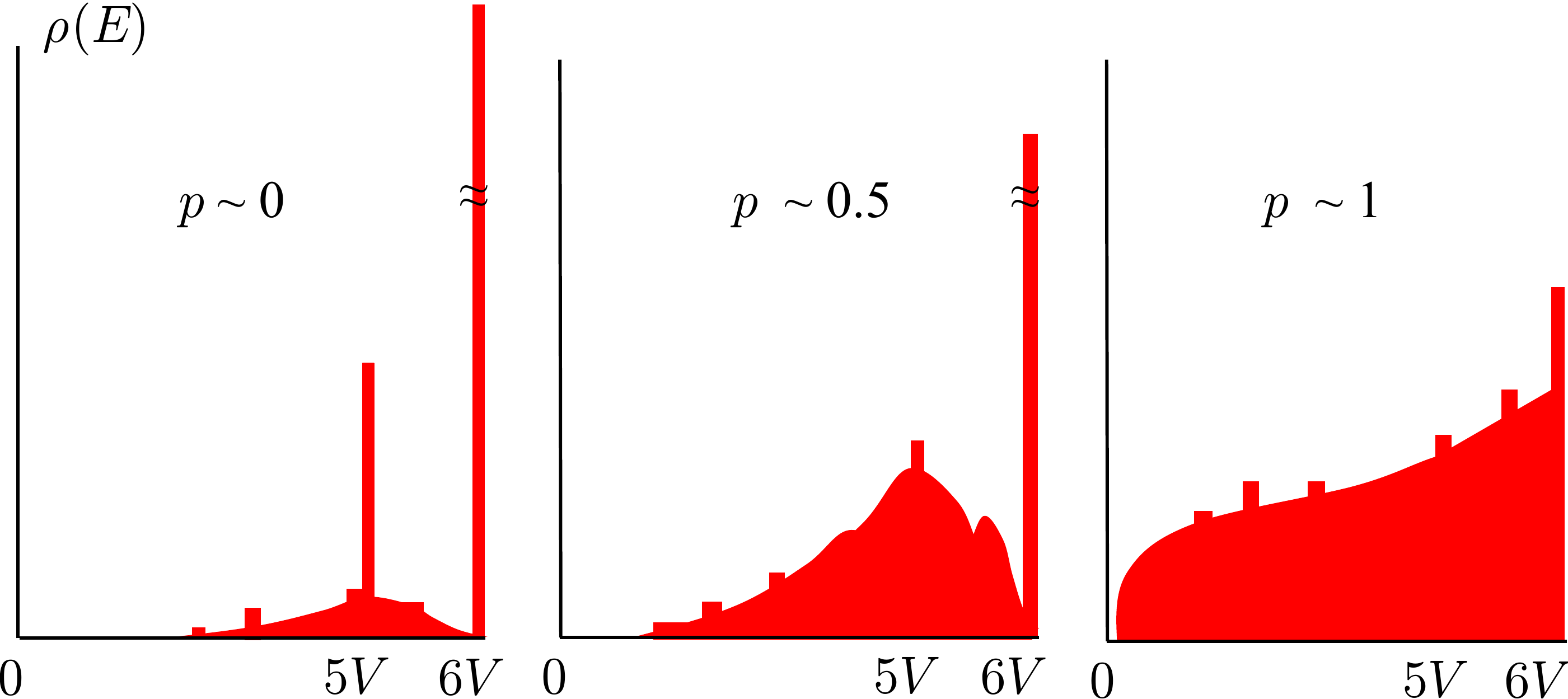}
  \caption{Schematic histograms of $\rho(E)$ for various baryon
    configurations as a function of eigenenergy $E$.  For dilute
    systems with $p\sim 0$, isolated single baryons are dominant and
    states with $E=\varepsilon_{\rm on}=6V$ are found the most
    frequently.  For larger $p$ the baryon cluster size grows up and
    states at smaller $E$ with larger spatial extension develop.}
  \label{fig:histogram}
\end{figure}

Shown in Fig.~\ref{fig:histogram} are schematic histograms of quark
eigenstates for various baryon configurations, namely, the density of
states $\rho(E)$ as a function of the eigenenergy $E$.  We note that a
sum rule, $\int dE \rho(E) =1$, should hold for a single particle
state.  In the dilute regime at $p\sim 0$ most baryons are isolated,
and eigenstates with the energy $E\simeq \varepsilon_{\rm on}=6V$ are
dominant.  The first nontrivial baryon configuration is a two-baryon
cluster for which the eigenenergy is
$\varepsilon_{\rm on}\pm V = 6V\pm V$.  Hence, the histogram is
expected to have peaks around $5V$ and $7V$ (in
Fig.~\ref{fig:histogram} the $E > 6V$ region is not shown as it should
be symmetric from the reference at $E=6V$).  As $p$ increases (i.e.,
the baryon density increases), configurations with isolated baryons
would be less populated, but sub-clusters of wave-function are formed
within the classical baryon clusters.  Some of sub-clusters make
contributions to $E=6V$ and the peak at $E=6V$ should persist up to
$p\sim 1$.

As long as sub-clusters appear, quarks are still localized, even
though they could flow from one baryon to another.  Therefore, Soft
Deconfinement is defined as complete delocalization of the quark
wave-function.  For $\varepsilon_{\rm off} \to \infty$ as is the case
here, it has been conjectured that localized states should appear at
$E=6V$ until the concentration reaches $p\to 1$.  In contrast, for
$\varepsilon_{\rm off}<\infty$ that allows for quantum tunneling, the
$E=6V$ state could be completely delocalized at $p<1$.

When the baryon density or $p$ gets larger, the typical spatial
extension of baryon clusters and thus quark wave-functions would be
larger.  This means that the lowest eigenenergy can be lowered and
softer quark components would be involved at larger baryon density.
Eventually, at the critical value of density or $p$, the wave-function
is delocalized in an arbitrary direction.  The eigenenergy reaches
$E=0$ when quarks are delocalized in all directions, and this is
possible only for $p=1$, as states with $p<1$ are accompanied by
vacancy of baryon clusters and it would lift up $k_i$ from zero and
thus $E>0$ inevitably.

\begin{figure}
  \includegraphics[width=0.7\columnwidth]{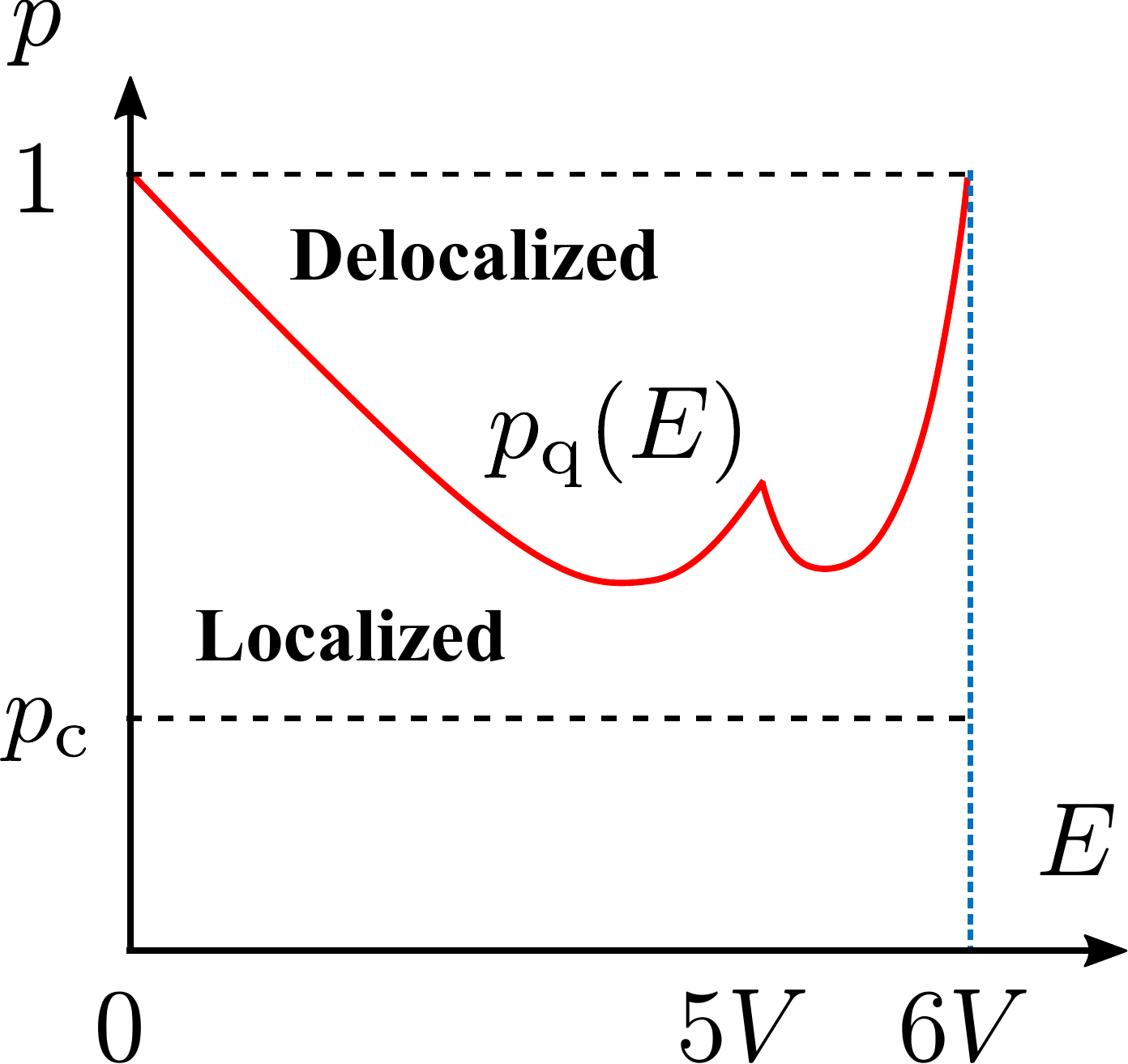}
  \caption{Schematic phase diagram of quantum percolation;
    the quantum critical concentration $\pq$ as a function of the
    eigenenergy $E$ and the classical critical concentration
    $\pc\;(<\pq)$ that is independent of $E$.  The minimum is around
    $E/V\sim \calO(1)$.  Cusps may appear corresponding to the
    molecular states.  The $E=0$ state would be realized only when
    baryons occupy all the sites, and thus $\pq(E\to 0)\to 1$.
    The $E=6V$ states are localized for any $p<1$ since single
    particle states get localized as long as finite cluster boundaries
    remain in the system.}
    \label{fig:qphase}
\end{figure}

Figure~\ref{fig:qphase} shows a schematic phase diagram of
percolation (for recent numerical studies, for example, see Fig.~3 in
Ref.~\cite{doi:10.7566/JPSJ.86.113704});  the critical concentration
of quantum percolation, $\pq(E)$, as a function of the eigenenergy $E$
and its classical counterpart, $\pc$, which is independent of $E$ but
solely determined by the geometrical site-bond networks.  For a given
density or $p$, modes with $p < \pq(E)$ are localized (they may be
extended over a wider range than a single baryon, but not delocalized
over the whole system), while modes with $p > \pq(E)$ can be
delocalized and deconfined.  As we mentioned before, the modes at
$E=0$ and $6V$ are somewhat special and can get delocalized only at
$p \to 1$.  The behavior of $\pq(E)$ in the small $E$ region is called
the \textit{mobility edge trajectory} and a relation between $\pq(E)$
and $\rho(E)$ are known~\cite{PhysRevB.36.8649}.  The minimum plateau
of $\pq(E)$ away from $E=0,\,6V$ reads $p_q(E) \gtrsim 1.3\pc$
typically.  Therefore, if we adopt Eq.~\eqref{eq:fillingsoft} for the
critical density $\sim 1.4\rho_0$ in the classical percolation
picture, the quantum effects would raise it up to
$\gtrsim 1.8\rho_0$.

Here, let us summarize our considerations based on the quantum
percolation model in the context of nuclear and quark matter.
Actually, the quantum percolation model provides us with useful
insights as follows.

First, the histograms of $\rho(E)$ as in Fig.~\ref{fig:histogram}
quantify how the quark eigenstates change as baryon clusters merge at
various baryon densities.  This way of understanding matter implicitly
assumes duality between baryons and quarks.  The point is that
$\rho(E)$ carries information on quarks for such many-body systems of
baryons.  As the density increases, the average kinetic energy of
baryons should increase, and at the same time, a larger baryon cluster
would allow for quarks with smaller average kinetic energies.  For
Soft Deconfinement quark momentum eigenstates would form natural bases
to characterize the nature of localization/delocalization of quarks.
It is likely that the changes in $\rho(E)$ occur continuously with
increasing density, so that the quantum percolation takes place
mode-by-mode gradually.  This is a microscopic description of the
quark-hadron continuity (apart from symmetries and gap energies).
Because $\rho(E)$ is positive definite, the gauge average would not
wash $\rho(E)$ out, and a gauge-invariant characterization would be in
principle possible.

Second, localization over baryon clusters can be driven by quantum
interference effects.  This observation happens to be consistent with
a conventional view of quark confinement.  In QCD it has been
established that the strong coupling limit of the theory should
confine quarks due to randomness of gluons.  In the present study we
saw that simple configurations as in Fig.~\ref{fig:examples_clusters}
exhibit rich contents in physics at the quantum level.  There are
various sources for randomness on top of baryon configurations, and it
would deserve more investigations whether gluon fluctuations
strengthen/weaken our proposed scenario.  This question is beyond our
current scope, and we just mention possibly related preceding works,
Refs.~\cite{GarciaGarcia:2005vj,Giordano:2016vhx,Holicki:2018sms},
in which chiral symmetry breaking and Anderson localization have been
discussed.

\subsection{Clustering and delocalization} \label{clustering}

\begin{figure*}
  \includegraphics[width=0.8\textwidth]{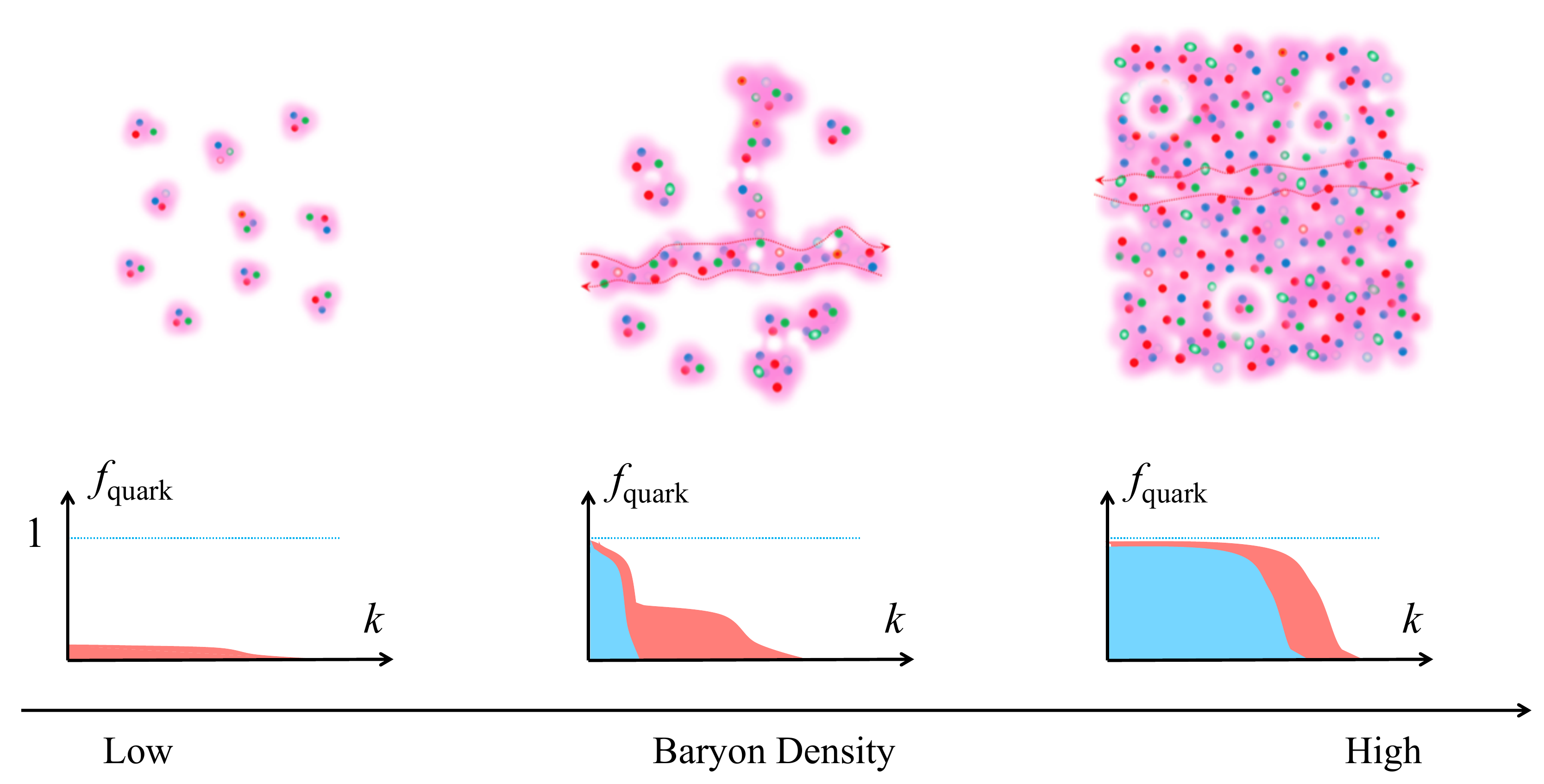}
  \caption{Graphical representation of Soft and Hard Deconfinement
    based on the percolation picture.  The occupation function,
    $f_{\rm quark} (k)$, for quarks with momenta $k$ is also
    schematically illustrated.  The red (blue) area in
    $f_{\rm quark}(k)$ indicates the contributions from localized
    (delocalized) modes.}
    \label{fig:percolation}
\end{figure*}

Finally we mention a possibility of the quantum percolation picture to
give us a clue to understand Quarkyonic Matter~\cite{McLerran:2007qj}
better from the nuclear point of view.  Quarkyonic Matter is well
defined only in the limit of $\Nc\to\infty$ in which gluons are
unscreened due to $1/\Nc$ suppression of quark loops and quarks are
still confined.  Then, one may encounter a conceptual question;  the
baryon Fermi sea or the quark Fermi sea, which of them should be a
more suitable starting point for quantitative estimates.  The presence
of confining forces via unscreened gluons would suggest the baryonic
description, but the pressure turns out to be $\calO(\Nc)$, and one
should account for quark degrees of freedom while keeping track of the
identity of baryons.  For this reason with two seemingly conflicting
aspects, it has been argued that the Fermi sea should be composed of
quarks, and any excitations on top of it should be confined.

It is indeed possible to form a color-singlet Fermi sea by filling all
colored states with quarks.  The Quarkyonic Matter scenario presumes
$\Nc$-particle correlations near the Fermi surface which form a
color-singlet, i.e., a baryonic composite.  Altogether, this picture
leads to a model of the momentum space
shell~\cite{McLerran:2018hbz,Jeong:2019lhv}.  However, it seems
counter-intuitive to postulate quarks at lower momenta and baryons at
larger momenta.

Now, let us discuss such a possible momentum shell in our language of
the mode-by-mode deconfinement.  The schematic illustration is shown
in Fig.~\ref{fig:percolation} where the occupation function,
$f_{\rm quark} (k)$, for quarks with momenta $k$ is also sketched.  In
the dilute regime with $\rho \ll \rsoft^{-3}$, the isolated baryons
are dominant as shown in the far left panel of
Fig.~\ref{fig:percolation}.  The quark momentum distribution is
characterized merely by quark compositions inside each baryon.
Therefore, $f_{\rm quark}(k)$ should have a support up to
$k\sim \rsoft^{-1} \sim \Lambda_{\rm QCD}$.  With increasing baryon
density, as long as baryons are isolated, $f_{\rm quark}(k)$ is simply
piled up without changing its shape itself.

As the baryon density increases further, baryons can be clustered, as
confirmed, for example, in numerical simulations of quantum molecular
dynamics~\cite{Maruyama:1997rp}.  This happens at the density,
$\rho \sim \rsoft^{-3}$, parametrically.  Nucleons are still
distributed dilutely at the normal nuclear density, but eventually,
the classical and the quantum percolation of baryon clusters would be
realized beyond a certain critical density of Soft Deconfinement, as
illustrated in the middle panel of Fig.~\ref{fig:percolation}.  On
such percolated clusters of baryons, the quark wave-functions can be
delocalized with low momentum components.  Then, these delocalized
states would substantially contribute to $f_{\rm quark}(k)$ in small
$k$ regions as indicated by blue area in the middle panel of
Fig.~\ref{fig:percolation}.  Meanwhile, localized states associated
with isolated baryons and small-sized baryon clusters still make
nonzero contributions to $f_{\rm quark}(k)$, but percolating quarks
emerge from the small-$k$ regions as the cluster domains get larger.
It should be noted that we do not necessarily assume inhomogeneous
baryonic states.  We should recall that our picture is based on the
Born-Oppenheimer approximation, and the true ground state properties
are obtained after taking the average over baryon configurations.  The
uniformity is recovered through the averaging procedure in the end.

At even larger densities where baryon hard cores overlap,
$\rho \sim \rhard^{-3}$, most clusters get quantum percolated, and
delocalized quark states become more populated as shown in the far
right panel of Fig.~\ref{fig:percolation}.  
In this way the quark Fermi sea grows up and develops with low momentum states of quarks
saturated.  Isolated baryon clusters would become fewer, although they
should still remain due to quantum interference effects.  Possibly,
therefore, $f_{\rm quark}(k)$ sustains contributions from localized
(interpreted as confined) states, as sketched by the red area in
$f_{\rm quark}(k)$ in the far right panel of Fig.~\ref{fig:percolation}.
These localized states on top of the Fermi sea may be regarded as relativistic baryons with $\Nc$-quarks
collectively moving in the same direction. 
This is in contrast to the case without the Fermi sea, 
where moving directions of $\Nc$-quarks are not aligned, leaving a small baryon momentum.
Changes from the non-relativistic to relativistic regime
can be one of sources for stiffening in equations of state~\cite{McLerran:2018hbz,Jeong:2019lhv}.

The modeling of dense matter as a superposition of localized and
delocalized quark wave-functions may be relevant to physical
observables involving excited modes rather than bulk thermodynamics.
Thus, the effects on the neutron star EoS may be limited, but the
transport coefficients such as the heat conductivity, the baryon
number diffusion constant, the viscosity, and so on could be sensitive
to details of mode-by-mode localization/delocalization.  
Another quantity sensitive to excitation modes
is the entropy density for which nuclear matter and quark matter
contribute in parametrically distinct ways.
The Soft-Deconfinement contains both of these contributions as in
Fig.~12, and it may be possible to define an effective critical
density separating nuclear, Soft-Deconfinement, and Hard-Deconfinement
regimes, in a way similar to the pseudo-critical temperature in finite
temperature QCD.{}  In principle these effects can be studied from
phenomenology such as protoneutron stars and neutron star mergers.
%
The photo production rate may be also an interesting indicator in a similar
fashion to high-$T$ matter coupled with the Polyakov
loop~\cite{Satow:2015oha}.

\section{Conclusions}
\label{sec:conclusions}

In this work we proposed two characterizations of quark deconfinement,
namely, Hard Deconfinement and Soft Deconfinement.  It is conceptually
straightforward to understand Hard Deconfinement along the lines of a
conventional picture of classical percolation of baryons.  Once the
nucleon core regions overlap, thermodynamic properties are dominated
by the energy-momentum tensor in the nucleon core that could be
available by measuring the gravitational form factors of the nucleon
in deeply virtual Compton scattering.  Based on this speculation, we
quantified the internal structures of a nucleon using a chiral soliton
model, and estimated the equation-of-state of its compact core.  We
found that the nucleon EoS obtained in this way is fairly consistent
with the empirical EoSs known from neutron star phenomenology.  We
also discussed implications of partial chiral symmetry restoration in
dense matter which lead to a stiffer EoS, a direction bringing the
nucleon core EoS closer to other empirical EoSs.

Soft Deconfinement is a more subtle notion.  Microscopically,
expectation values are taken over ensembles of various quantum
states.  Consider various snapshots of baryon configurations.  As the
baryon density increases, the cluster size in such snapshots of baryon
configurations becomes larger.  Then, eventually, quark wave-functions
are more and more delocalized and the quantum percolation of quarks
occurs from smaller momentum modes of quarks.

The appropriate physical interpretation of Soft Deconfinement should
thus be based on the localization or delocalization of quark states.
Even if they are localized, quark wave-functions may extend over a
wider range than a single nucleon, and so there is no sharp
identification of confinement/deconfinement.  Such gradual changes of
the delocalization range would give us detailed insights on the
quark-hadron continuity, and at the same time, useful clues to resolve
intuitive views of Quarkyonic Matter and its field-theoretical
modeling (see also Ref.~\cite{Cao:2020byn} for a recent attempt).  In
contrast to Hard Deconfinement for which we presented quantitative
estimates, our discussions on Soft Deconfinement are limited to a
qualitative level.  It is a very interesting and challenging problem
to formulate a quantitative description of $\rho(E)$ and reveal its
gauge (in)dependence, so that the mobility edge trajectory can be
drawn and the detailed momentum shell structures could be clarified.
\vspace*{0.8em}

\begin{acknowledgments}
  We thank
  Gordon~Baym,
  Yuki~Fujimoto,
  Hideaki~Iida,
  Giorgio~Torrieri,
  Naoki~Yamamoto for
  useful discussions and comments.
  K.~F.\ was supported by Japan Society for the Promotion
  of Science (JSPS) KAKENHI Grant Nos.\ 18H01211 and 19K21874.
  T.~K. was supported by NSFC Grant No.\ 11875144.
\end{acknowledgments}

\bibliographystyle{apsrev4-1}
\bibliography{softhard}

\end{document}